# Lilienfeld Transition Radiation Brought to Light

**Mario Rabinowitz**
*Electric Power Research Institute*
*Palo Alto, California 94303*

Inquiries to: *Armor Research*
*715 Lakemead Way, Redwood City, CA 94062*
Mario715@earthlink.net

**Abstract**

Presently, the term transition radiation tends to denote a somewhat different physical phenomenon than the original transition radiation discovered by J. E. Lilienfeld in 1919, and re-employed in different forms again in 1953 and 1971. Lilienfeld transition radiation is a subtle kind of radiation with distinctive properties that may often be unexpected in prospect, yet is intuitive and readily understood in retrospect. This paper distinguishes in clear terms between the different kinds of transition radiation, and shows its link to modern apparatus such as possible applications in the area of x-ray microscopy, microholography, and the free-electron laser .

## I. Introduction

For a static charge, it is well known that the solution for the net field outside a flat metal surface is the superposition of the field of the charge and that of its mirror-image opposite charge in the metal. For a negative charge, this is due to the repulsion of the free electrons in the metal. For a positive charge this is due to the attraction of the free electrons. In more mathematical terms, the net field may be found by solving Laplace's equation with the appropriate boundary conditions to obtain the potential. The field is minus the gradient of the potential. The result is clearly an electric dipole.

Thus for a charge moving slowly, its field penetrates well into the metal and a simple field pattern results. For a very fast moving charge, the field only penetrates with an exponential decay so that the field is largely limited within a skin depth of the metal's surface. Easily tractable fields result when the distance to the surface of the charge is small compared with the penetration depth. As we shall see, in either case radiation results even when no external force is applied to accelerate the charge, if the metal surface is not smooth.

## II. Radiation from Uniform motion

The radiation described above is not the only effect in which radiation occurs from seemingly uniform motion. Cerenkov radiation also occurs with no acceleration of the charge, when its speed exceeds the speed of light in a given medium as first predicted by Arnold Sommerfeld.

Analogously, if a charge moves at constant velocity through a junction between two materials having different opitical properties, radiation results from the change in the field the charge is carrying. The term "Transition Radiation" is used to denote radiation emanating when charged particles go from one medium to another. The theory of this radiation was presented by Frank and Ginsburg [1] in 1944, and is not at all the same as Cerenkov radiation; or the next described Lilienfeld transition radiation.

## III. Comparing Different Kinds of Radiation

In 1919 Lilienfeld found that in addition to x rays, radiation ranging from visible light through the ultraviolet is emitted when electrons approach a metal electrode.[2] This radiation has a characteristic polarization, spectrum and intensity. Lilienfeld transition radiation can be considered to originate from the time rate of change of the virtual dipole between charged particles and their image charges that forms as the charged particles move near a conducting surface, as described in Section I. In Lilienfeld's original experiment, the charged particles were low-energy electrons moving toward a metallic anode. This was originally called, "Lilienfeld Transition Radiation." Largely because of high energy phyics applications Cerenkov Radiation and Transition Radiation are currently well known. Unfortunately, Lilienfeld Transition Radiation has been largely forgotten.

In a variation of Lilienfeld transition radiation, the charged particles move roughly parallel to a conducting serrated surface, producing an oscillating virtual dipole whose frequency is related to the particle velocity and the serration spacing. This 1953 work has been called "Smith-Purcell Radiation." [3] In 1992, Michael M. Moran [4] calculated the potential of x-ray generation by the Smith-Purcell effect. He concluded that "although the radiated power may be low, the source *brightness* could be surprisingly large because of the microscopic source area. This together with the tunability and inherent spatial coherence of S-P radiation, suggests possible applications in the area of x-ray microscopy or microholography." In 1992 Doucas et al [6] used a beam of 3.6 MeV electrons to generate far infrared radiation by the S-P effect. Until then low-energy electrons were used to produce visible radiation. Their "emphasis is on a spectral region where the population of sources is sparse."

In some ways Lilienfeld's work anticipated the 1953 experiment of S. J. Smith and Edward M. Purcell [3], and possibly even John M. J. Madey's invention of the free-electron laser. In 1971, John M. J. Madey [5] used the Weizsacker-Williams method to calculate the gain due to the induced emission of radiation by a relativistic motion of an electron moving through a periodic transverse grating. The magnetic grating may be considered analogous to the periodically serrated conducting surface of Smith and Purcell, with the important added bounus of laser gain. Each may be considered to be an important and novel variation of the prior work, as reflected in much of the progress of science.

As a graduate student I observed Lilienfeld transition radiation in 1961 in connection with my doctoral thesis research, before I had ever heard of it. A literature search led quickly to Lilienfeld's work. Visible polarized blue light is easily seen at the anode of a high voltage vacuum tube.